\documentclass[aps,showpacs,superscriptaddress,twocolumn,prb]{revtex4-1}
\usepackage{hyperref}
\usepackage{graphicx}
\usepackage{amsmath,amssymb,amsfonts}

\begin{document}

\title{Anisotropic magnetoresistance and upper critical fields up to 63 T in CaKFe$_4$As$_4$ single crystals}

\author{Tai Kong}  
\affiliation{Ames Laboratory, U.S. DOE, and Department of Physics and Astronomy, Iowa State University, Ames, Iowa 50011, USA}

\author{Fedor F. Balakirev}
\affiliation{National High Magnetic Field Laboratory, Los Alamos National Laboratory, MS-E536, Los Alamos, New Mexico 87545, USA}

\author{William R. Meier}  
\affiliation{Ames Laboratory, U.S. DOE, and Department of Physics and Astronomy, Iowa State University, Ames, Iowa 50011, USA}

\author{Sergey L. Bud'ko}  
\affiliation{Ames Laboratory, U.S. DOE, and Department of Physics and Astronomy, Iowa State University, Ames, Iowa 50011, USA}

\author{Alex Gurevich}
\affiliation{Department of Physics, Old Dominion University, Norfolk, Virginia 23529, USA}

\author{Paul C. Canfield}
\affiliation{Ames Laboratory, U.S. DOE, and Department of Physics and Astronomy, Iowa State University, Ames, Iowa 50011, USA}

\begin{abstract}

We report the temperature dependencies of the upper critical fields $H_{c\text{2}}^{\text{c}}(T)$ parallel to the c-axis and $H_{c\text{2}}^{\text{ab}}(T)$ parallel to the ab-plane of single crystalline CaKFe$_4$As$_4$ inferred from the measurements of the temperature-dependent resistance in static magnetic fields up to 14 T and magnetoresistance in pulsed fields up to 63 T. We show that the observed decrease of the anisotropy parameter $\gamma(T)=H_{c\text{2}}^{\text{ab}}/H_{c\text{2}}^{\text{c}}$ from $\simeq 2.5$ at $T_c$ to $\simeq 1.5$ at 25 K can be explained by interplay of paramagnetic pairbreaking and orbital effects in a multiband theory of $H_{c2}$. The slopes of $dH_{c\text{2}}^{\text{c}}/dT\simeq-4.4$ T/K and $dH_{c\text{2}}^{\text{ab}}/dT \simeq-10.9$ T/K at $T_c$ yield an electron mass anisotropy of $m_{ab}/m_c\simeq 1/6$ and short coherence lengths $\xi_c\simeq 5.8\,\text{\AA}$ and $\xi_{ab}\simeq 14.3\,\text{\AA}$. The behavior of $H_{c\text{2}}(T)$ turns out to be similar to that of the optimal doped (Ba,K)Fe$_2$As$_2$, with $H_{c\text{2}}^{\text{ab}}(0)$ extrapolating to $\simeq 92$ T, well above the BCS paramagnetic limit.

\end{abstract}

\maketitle

The discovery of Fe-based superconductors (FBS) \cite{Kamihara08} has intensified research on mechanisms of high-temperature-superconductivity and searches for materials with higher superconducting transition temperatures $T_c$. Among the many different classes of FBS\cite{Stewart11}, the so called "122" family with a ThCr$_2$Si$_2$-type structure (I4/mmm) is one of the most well-studied systems. Superconductivity in the "122" family was first discovered in (Ba,K)Fe$_2$As$_2$\cite{Rotter08b} with alkali metal (A) /alkali earth (Ae) substitution. Subsequently it was found that superconductivity can also be induced by transition metal substitutions\cite{Sefat08,Ni08Ba,Canfield10} or substitutions on the As site\cite{Jiang09}. In these substitution cases, though, the ThCr$_2$Si$_2$-type structure remained the same. However, when the difference in ionic radii between the A and Ae ions becomes larger, such as in the case of Ca and K, a homogeneous substitution on the A/Ae site could not be reached\cite{Wang13}.

Recently, Iyo et al.\cite{Iyo16} systematically studied the structure of (A,Ae)Fe$_2$As$_2$ polycrystals and stabilized a new AeAFe$_4$As$_4$ structure type (P4/mmm) with alternating A/Ae layers separated by Fe-As layers when the A/Ae ionic radii are sufficiently different. Unlike a homogeneous substitution within the ThCr$_2$Si$_2$-type structure, each A/Ae in the AeAFe$_4$As$_4$ structure occupies a unique, well-defined, crystallographic site. Single crystals of CaKFe$_4$As$_4$ were synthesized and characterized soon after the initial discovery\cite{Meier16}. CaKFe$_4$As$_4$ is found to be superconducting at $\sim$35 K with no other phase transitions from 1.8 K to 300 K\cite{Meier16}. The pressure dependence of $T_c$, anisotropic electric resistivity, elastoresistivity, Hall effect and thermal power data suggest that CaKFe$_4$As$_4$ behaves similarly to optimal, or slightly over-doped, (Ba,K)Fe$_2$As$_2$\cite{Meier16}. Anisotropic $H_{c\text{2}}$($T$) data were measured up to 14 T; over this very limited field and temperature range, the anisotropic $H_{c\text{2}}$($T$) curves for CaKFe$_4$As$_4$ look a lot like those of Ba$_{0.55}$K$_{0.45}$Fe$_2$As$_{2}$\cite{Altarawneh08}. But, clearly, significantly higher field data will be necessary to reveal $H_{c\text{2}}$ at low temperatures.

The newly discovered CaKFe$_4$As$_4$ is expected to have very high upper critical fields $H_{c\text{2}}(T)$ mediated by the characteristic of FBS interplay of multiband orbital effects and strong Pauli pairbreaking \cite{Gurevich10,Gurevich11}. Yet the extent to which the above features of CaKFe$_4$As$_4$ can manifest themselves in the high-field behavior of $H_{c\text{2}}(T)$ and whether it would be different from that of the well-studied 122 FBS \cite{Kano09,Tarantini11} has not yet been addressed. Another intriguing question is whether multiband superconductivity in ordered CaKFe$_4$As$_4$ could result in crossing of the upper critical field curves $H_{c\text{2}}^{\text{c}}(T)$ parallel to the c-axis and $H_{c\text{2}}^{\text{ab}}(T)$ parallel to ab-plane, as has been observed in some FBS and other superconductors\cite{Fang10,Maiorov14,Balakirev15}. The anticipated lack of the effect of disorder on multiband superconductivity in CaKFe$_4$As$_4$ also makes this material (along with other such stoichiometeric FBS as LiFeAs \cite{Cho11}) a good candidate for searching for the Fulde-Ferrel-Larkin-Ovchinnikov (FFLO) states \cite{Gurevich10,Gurevich11,Matsuda07,Lebed11,Croitoru12}. To address these points, in this paper we present measurements of anisotropic magnetoresistance and $H_{c\text{2}}$($T$) in CaKFe$_4$As$_4$ single crystals in high magnetic fields up to 63 T.

Samples were grown using a high-temperature solution growth technique out of excess FeAs. Elemental K (Alfa Aesar, 99.95$\%$), Ca (Ames Lab, 99.9+$\%$) and pre-reacted FeAs (Fe: Alfa Aesar, 99.9+$\%$. As: Alfa Aesar, 99.9999$\%$) with a starting molar ratio of Ca:K:FeAs = 0.8:1.2:10 were held in an alumina frit-disc crucible set\cite{Canfield16} and sealed in Ta tube, which were then sealed in a silica jacket under partial Ar atmosphere\cite{Kong15a}. Slightly more K than Ca was used to suppress the formation of CaFe$_{2}$As$_{2}$. The packed ampoule was then heated up to 1180 $^o$C, held at 1180 $^o$C for 5 hours, rapidly cooled to 1050 $^o$C and then slowly cooled to 930 $^o$C, at which temperature the single crystals and the liquid was separated in a centrifuge. Single crystals are plate-like with typical dimension of 5$\times$5$\times$0.1 mm. More detailed growth procedures can be found in Ref.~\onlinecite{Meier16}.

The upper critical field $H_{c\text{2}}$($T$) was inferred from transport magnetoresistance measured by a standard four-probe technique. Both DuPont 4929N silver paint and Epotek-H20E silver epoxy were used to attach contact leads onto the samples (Pt for measurements static field measurements and twisted copper wires for pulsed field measurements). For static fields below 14 T, resistance was measured using a Quantum Design (QD) physical property measurement system, PPMS-14 ($T$ = 1.8-305 K, $H$ = 0-14 T, $f$ = 17 Hz.). Higher field data were obtained in a 63 T pulsed magnet at the National High Magnetic Field Laboratory (NHMFL), Los Alamos, using a high-frequency, synchronous digital lock-in technique ($f$ = 148 kHz).

\begin{figure}[!h]
\centering
\includegraphics[scale=0.31]{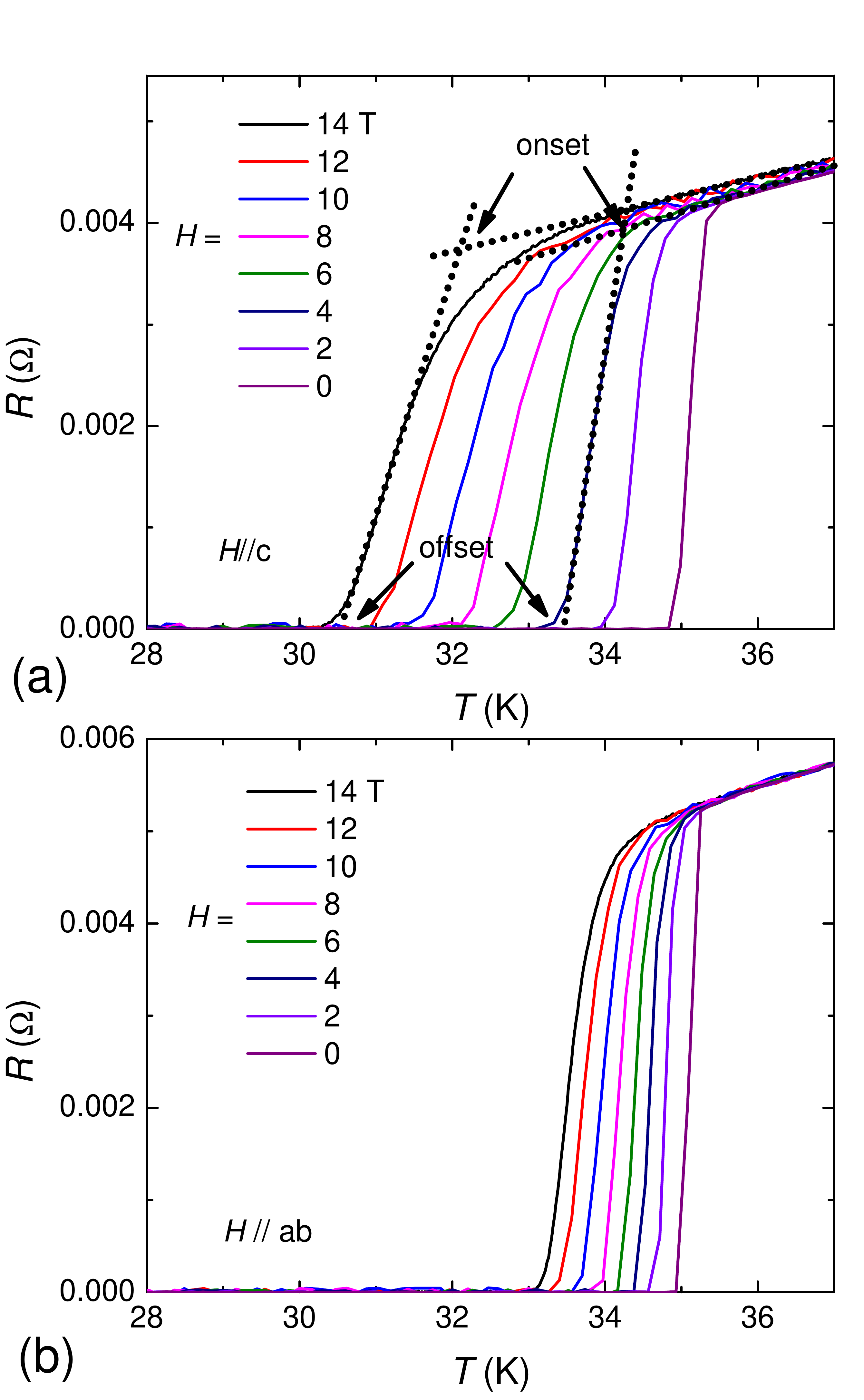}
\caption{(Color online) Temperature-dependent resistance measured in a QD PPMS at different applied fields with (a) $H\parallel$c and (b) $H\parallel$ab. Dotted line and arrows indicate different criteria for determining $H_{c\text{2}}$ (see text).}
\label{RT}
\end{figure}

Fig.~\ref{RT} shows the temperature-dependent resistance measured at different applied fields in a QD PPMS. The criteria for determining $T_{\text{c}}$ are shown in Fig.~\ref{RT}(a). The onset criterion identifies the temperature at which the normal state line intersects with the maximum slope of the resistance curve. The offset criterion identifies the temperature of the intersection of the maximum slope of the resistance curve and the zero-resistance line. In zero field, the superconducting transition is very sharp. As field is increased, the transition becomes slightly broader, more so for $H\parallel$c than for $H\parallel$ab. For $H\parallel$c, the T$_{\text{c}}$ value is suppressed to a lower temperature than for $H\parallel$ab at a given field. 

\begin{figure}[!h]
\centering
\includegraphics[scale=0.33]{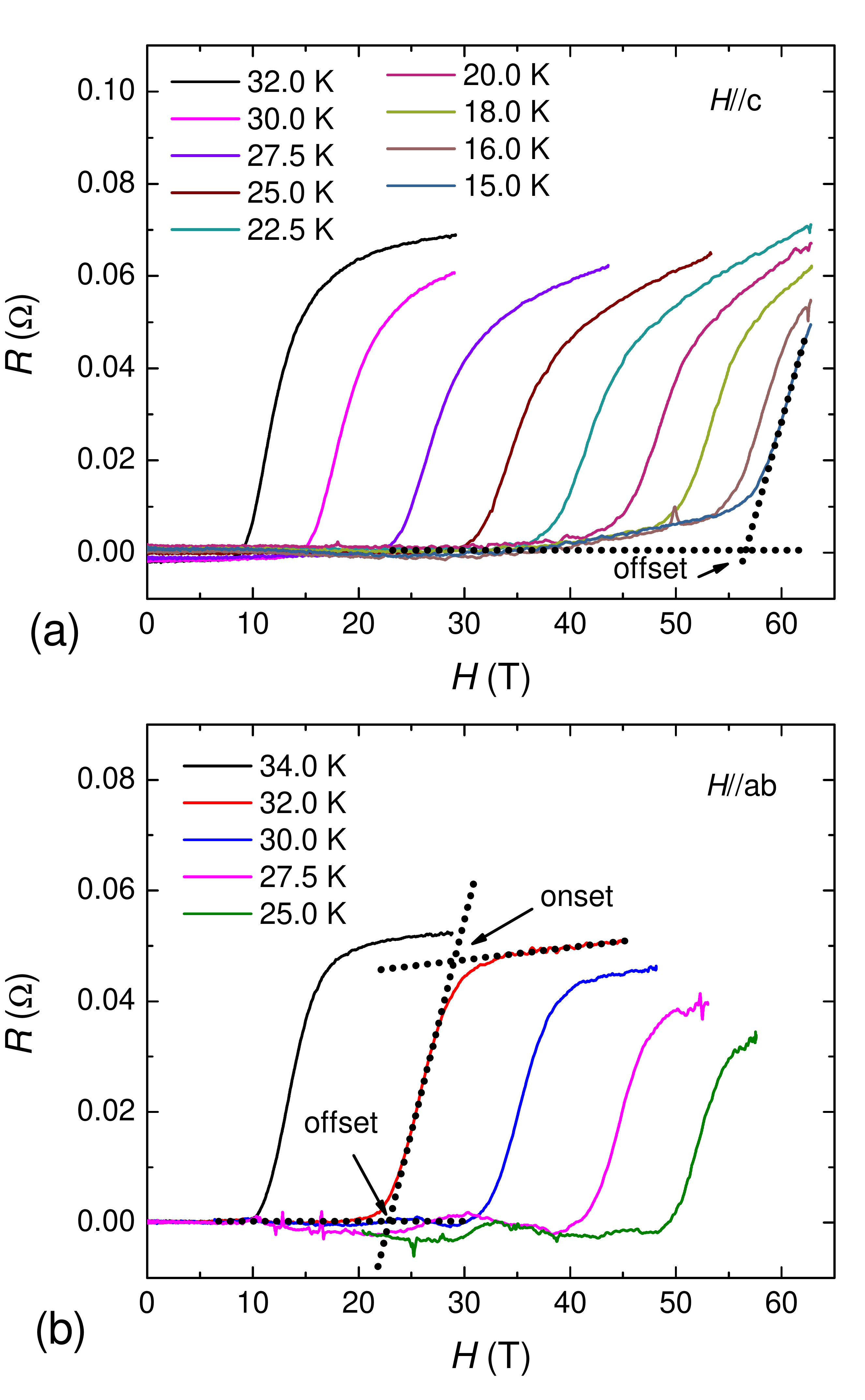}
\caption{(Color online) Field-dependent resistance measured in a 63 T pulsed magnet at different temperatures with (a) $H\parallel$c and (b) $H\parallel$ab. A temperature-independent background signal was subtracted for clarity (see text). Dotted line and arrows indicate different criteria for determining $H_{c\text{2}}$ (see text).}
\label{RH}
\end{figure}

Fig.~\ref{RH} shows the field-dependent resistance measured at different temperatures. A temperature-independent background was subtracted from the signal for clarity. The background is attributed to the displacement of the sample and its wiring by Lorentz force synchronous with lock-in excitation current. The resulting magnetic inductance voltage is a product of field intensity and Lorentz force, leading to a stray background signal proportional to $H^2$. Similar onset and offset criteria were applied to extract the superconducting field values at a given temperature. For $H\parallel$c at 15 K, only an offset value could be resolved as shown in Fig.~\ref{RH}(a).

\begin{figure}[!h]
\centering
\includegraphics[scale=0.35]{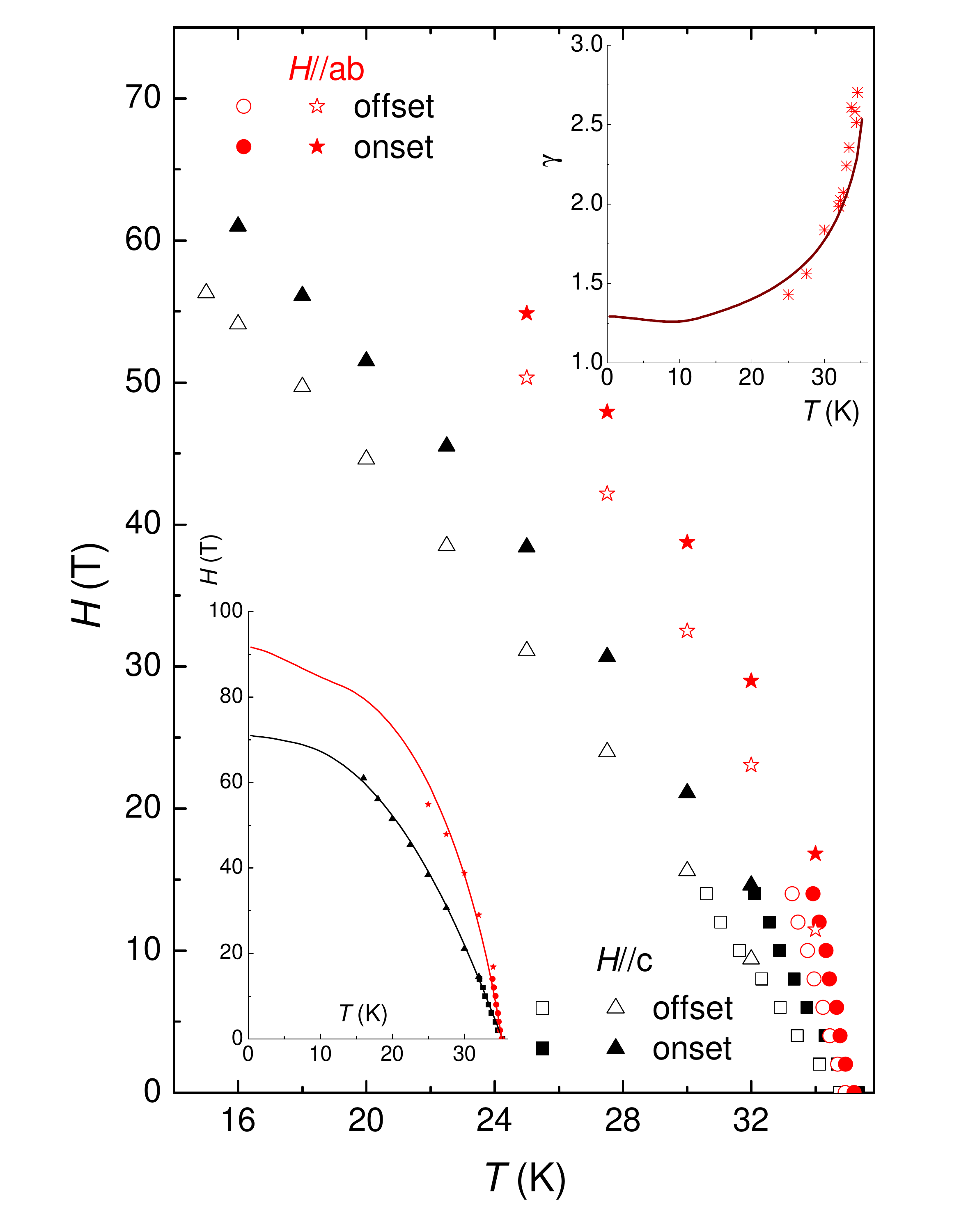}
\caption{(Color online) Anisotropic $H_{c\text{2}}$($T$) up to 61 T in CaKFe$_4$As$_4$ single crystals. Black squares (from PPMS) and black triangles (from pulsed magnet) represents $H_{c\text{2}}^{\text{c}}$($T$). Red circles (from PPMS) and red stars (from pulsed magnet) represents $H_{c\text{2}}^{\text{ab}}$($T$). Open and filled symbols indicate onset and offset criteria as described in the text. The upper inset shows the anisotropic parameter $\gamma(T)=H_{c\text{2}}^{\text{ab}}/H_{c\text{2}}^{\text{c}}$ together with the theoretically fitted curve (brown solid line). The lower inset shows the anisotropic $H_{c\text{2}}$($T$) data points from the onset criteria with the theoretically fitted curve (black and red solid lines). Previously, data measured by PPMS were presented in Ref.\onlinecite{Meier16}.}
\label{Hc2}
\end{figure}

The anisotropic $H_{c\text{2}}$($T$) data inferred from the temperature-dependent and field-dependent resistance data, summarized in Fig.~\ref{Hc2}, reveal multiple features about CaKFe$_4$As$_4$. (1) The values of $H_{c\text{2}}(0)$ both parallel and perpendicular to the $c$-axis extrapolate to the fields well above the single-band BCS paramagnetic limit $H_p [T] = 1.84T_c [K] \simeq 64$ T, which is close to the maximum field in our pulse magnet. Thus, Pauli pairbreaking is essential, similar to the majority of other FBS \cite{Gurevich11}. (2) As a result of different temperature dependencies of $H_{c\text{2}}^{\text{c}}(T)$ and $H_{c\text{2}}^{\text{ab}}(T)$ the anisotropy parameter $\gamma(T)=H_{c\text{2}}^{\text{ab}}(T)/H_{c\text{2}}^{\text{c}}(T)$ decreases as $T$ decreases (see upper inset in Fig.~\ref{Hc2}), consistent with the interplay of orbital and Pauli pairbreaking \cite{Gurevich10}. (3) No crossing of $H_{c\text{2}}^{\text{c}}(T)$ and $H_{c\text{2}}^{\text{ab}}(T)$ was observed at $0<H<61$ T, although a possibility that it may happen at higher fields cannot be ruled out.  

The superconducting coherence lengths $\xi_{ab}(T)=\xi_{ab}\tau^{-1/2}$ in the $ab$ plane and $\xi_c(T)=\xi_c\tau^{-1/2}$ along the $c$-axis, as well as the electron band mass anisotropy can be estimated from the measured slopes of $dH_{c\text{2}}^{\text{c}}/dT\approx -4.4$ T/K and $dH_{c\text{2}}^{\text{ab}}/dT\approx -10.9$ T/K at $T_c$ using the single-band Ginzburg-Landau (GL) theory, where $\tau=1-T/T_c$ [\onlinecite{Tinkham}]. From $|dH_{c\text{2}}^{\text{c}}/dT|=\phi_0/2\pi\xi_{ab}^2T_c$ and $|dH_{c\text{2}}^{\text{ab}}/dT|=\phi_0/2\pi\xi_{ab}\xi_cT_c$, where $\phi_0$ is the magnetic flux quantum, we obtain $\xi_{ab}\simeq 14.3\, \text{\AA}$ and $\xi_c\simeq 5.8\, \text{\AA}$. These coherence lengths are of the order of the lattice parameters\cite{Iyo16}: $a = 3.866 \, \text{\AA}$ and $c = 12.817 \,\text{\AA}$, the value $\xi_c$ being half the unit cell height along the c-axis. The resulting electron effective mass anisotropy $\epsilon = m_{ab}/m_c = (\xi_c/\xi_{ab})^2\simeq 1/6$ is similar to that of the 122 family of FBS and smaller than $m_c/m_{ab}\simeq 20-30$ characteristic of the 1111 FBS and YBa$_2$Cu$_3$O$_{7-x}$\cite{Gurevich10,Gurevich11}. Such short coherence lengths and the lack of structural disorder also suggest that CaKFe$_4$As$_4$ is in the clean limit as the mean free path $\ell \gg \xi_{ab}$ for the measured resistivity $\rho_n\sim 100\, \mu\Omega$ cm.

To gain further insight into the behavior of $H_{c\text{2}}(T)$, we fitted the experimental data using a two-band theory which takes into account both orbital and Pauli pairbreaking in the clean limit for two ellipsoidal Fermi surfaces. In this case the equation for $H_{c\text{2}}^{\text{c}}$ is given by \cite{Gurevich10,Gurevich11},
\begin{gather}
    a_1G_1+a_2G_2+G_1G_2=0,
\label{mgh} \\
    G_1=\ln t+2e^{q^{2}}\operatorname{Re}\sum_{n=0}^{\infty}\int_{q}^{\infty}due^{-u^{2}}\times
\nonumber \\
    \left[\frac{u}{n+1/2}-\frac{t}{\sqrt{b}}\tan^{-1}\left(
    \frac{u\sqrt{b}}{t(n+1/2)+i\alpha b}\right)\right].
\label{U1}
\end{gather}
Here $a_1=(\lambda_0 +\lambda_{-})/2w$, $a_2=(\lambda_0-\lambda_{-})/2w$, $\lambda_{-}=\lambda_{11}-\lambda_{22}$, $\lambda_0=(\lambda_{-}^2+4\lambda_{12}\lambda_{21})^{1/2}$, $w=\lambda_{11}\lambda_{22}-\lambda_{12}\lambda_{21}$, $t=T/T_c$, $\lambda_{11}$ and $\lambda_{22}$ are the pairing constants in band 1 and 2, and $\lambda_{12}$ and $\lambda_{21}$ are interband pairing constants. The function $G_2$ 
is obtained by replacing $\sqrt{b}\to\sqrt{\eta b}$ and $q\to q\sqrt{s}$ in $G_1$, where
\begin{gather}
    b=\frac{\hbar^{2}v^{2}_1 H_{c2} }{8\pi\phi_{0}k_B^2T_{c}^{2}},\qquad\alpha=\frac{4\mu\phi_{0}k_BT_{c}}{\hbar^{2}v^2_1},
    \label{parm1} \\
    q^{2}=Q^{2}\phi_{0}\epsilon_1/2\pi H_{c2}, \qquad \eta = v_2^2/v_1^2, \qquad
    s=\epsilon_2/\epsilon_1.
    \label{parm2}
\end{gather}
Here $Q$ is the wave vector of the FFLO modulations of the order parameter, $v_j$ is the in-plane Fermi velocity in band $j={1,2}$, $\epsilon_{j}=m_{j}^{ab}/m_{j}^c$ is the mass anisotropy ratio, $\mu$ is the magnetic moment of a quasiparticle, $\alpha\approx 1.8\alpha_M$, and $\alpha_M=H_{c\text{2}}^{orb}/\sqrt{2}H_p$ is the Maki paramagnetic parameter. If $\textbf{H}$ is applied along the symmetry axis, $\textbf{Q}$ is parallel to $\textbf{H}$ and the magnitude of $Q$ is determined by the condition $\partial H_{c\text{2}}/\partial Q=0$ of maximum $H_{c\text{2}}$.

 For the sake of simplicity, we consider here the case of $\epsilon_1=\epsilon_2=\epsilon$ for which the anisotropic $H_{c\text{2}}$ can be written in the scaling form 
 $$
 H_{c\text{2}}^{\text{c}}(T) = H_0b(t,\eta,\alpha),\quad H_{c\text{2}}^{\text{ab}}(T) = \frac{H_0}{\sqrt{\epsilon}}b\left(t,\eta,\frac{\alpha}{\sqrt{\epsilon}}\right), 
 $$
 where $H_0=8\pi\phi_0k_B^2T_c^2/\hbar^2v_1^2$ and $b$ is a solution of Eq. (\ref{mgh}). The fit of the measured $H_{c\text{2}}(T)$ to Eq. (\ref{mgh}) for $s_{\pm}$ pairing with $\lambda_{11}=\lambda_{22}=0$, $\lambda_{12}\lambda_{21}=0.25$, $\eta = 0.2$, $\alpha=0.5$, and $\epsilon = 1/6$ is shown in Fig.~\ref{Hc2} where $H_0$ was adjusted to fit the magnitude of $H_{c\text{2}}^{\text{c}}(T)$. The value of $\alpha$ is consistent with those which have been used previously to describe $H_{c\text{2}}(T)$ of Ba$_{1-x}$K$_x$As$_2$Fe$_2$ \cite{Tarantini11}.  
 
The fit shows that the upper critical fields at $T=0$ extrapolate to $H_{c\text{2}}^{\text{c}}(0) \approx 71$ T and $H_{c\text{2}}^{\text{ab}}(0) \approx 92$ T, the shape of $H_{c\text{2}}^{\text{c}}(T)$ being mostly determined by orbital effects moderately affected by the Pauli pairbreaking. By contrast, the shape of $H_{c\text{2}}^{\text{ab}}(T)$ is consistent with the essential Pauli pairbreaking in both bands, because of large respective Maki parameters $\alpha_1^{ab} = \alpha/\sqrt{\epsilon}$ and $\alpha_2^{ab} = \alpha/\eta\sqrt{\epsilon}$. In the available field range $0<H<61$ T where the $H_{c2}$ data were obtained, the fit is not very sensitive to the particular values of the pairing constants and the band asymmetry parameter $\eta$, yet it suggests the FFLO state at $T < 13$ K and for higher fields $H$ parallel to the $ab$ planes. More definite conclusions about multiband orbital effects and FFLO states could be made by analyzing low-temperature parts of the $H_{c\text{2}}^{\text{c}}(T)$ and $H_{c\text{2}}^{\text{ab}}(T)$, which would require even higher fields $H>63$ T. This distinguishes CaKFe$_4$As$_4$ from other ordered stoichiometric FBS compounds like LiFeAs for which the entire anisotropic $H_{c\text{2}}(T)$ has been measured \cite{Cho11}.   

Further insights into the magneto-transport behavior of CaKFe$_4$As$_4$ can be inferred from the fact that the resistance transition curves $R(T)$ shown in Figs. 1 and 2 broaden as $H$ increases. This indicates an essential effect of thermal fluctuations of vortices similar to that has been extensively studied in high-$T_c$ cuprates in which a superconductor at fields not far below $H_{c2}(T)$ is in a thermally-activated flux flow described by the ohmic resistivity\cite{Blatter94}:
\begin{equation}
\rho(T)=\rho_0 e^{-U(T,H)/k_BT},
\label{taff}
\end{equation} 
where $U(T,H)=U_0 (1-T/T_c)^\alpha (H_{c2}(0)/H)^{\beta}$ is the activation barrier, $U_0$ and the exponents $\alpha$ and $\beta$ depend on details of pinning, and $\rho_0\sim \rho_n$ \cite{Blatter94}. 
If the offset temperature $T_m$ in Fig. 1 is defined at a resistivity $\rho_m = \rho(T_m) \ll \rho_n$, it follows from Eq. (\ref{taff}) that the width of the resistive transition $T_c-T_m$ increases with $H$:
\begin{equation}
T_c - T_m \simeq T_c \left[\frac{k_BT_c}{U_0}\ln\frac{\rho_n}{\rho_m}\right]^{1/\alpha}\left[\frac{H}{H_{c2}(0)}\right]^{\alpha/\beta}
\label{tmm}
\end{equation}   
Broadening of the superconducting transition in CaKFe$_4$As$_4$ under magnetic field was also observed by measuring the step in the specific heat\cite{Meier16}. 

At $H=0$ thermal fluctuations can be quantified by the Ginzburg number $Gi=0.5(2\pi \mu_0 k_BT_c\lambda_0^2/\phi_0\xi_c)^2$ expressed in terms of $\xi_c$ and the London penetration depth $\lambda_0$ 
at $H||c$ and $T=0$. The values of $\lambda_0$ in CaKFe$_4$As$_4$ has not yet been measured, but for $\lambda_0=100-200$ nm characteristic of the majority of FBS, we obtain that CaKFe$_4$As$_4$ with $\xi_c=0.6$ nm and $T_c=35$ K, would have $Gi=1.25\cdot 10^{-4}-2\cdot 10^{-3}$, of the same order of magnitude as $Gi$ for other FBS but smaller that $Gi\sim 10^{-2}$ for YBa$_2$Cu$_3$O$_{7-x}$ \cite{Putti10,Gurevich14}. The offset point of $R(T,H)=0$ in Fig. 1 can also be associated with the irreversibility field $H_p(T)$ of melting and thermal depinning of vortex structure. For instance, the melting field $H_m$ of the ideal vortex lattice in a uniaxial superconductor at $H\|c$ is defined by the equation $h_m/(1-h_m)^3=(1-t)t_0^2/t^2$, where $h_m=H_m/H_{c\text{2}}$, $t_0=\pi c_L^2/Gi^{1/2}$ and $c_L=0.15-0.17$ is the Lindemann number \cite{Blatter94}. For weak thermal fluctuations, $H_{c\text{2}}-H_m \ll H_{c\text{2}}$, the above equation for $h_m$ yields
\begin{equation}
H_{c\text{2}}(T)-H_m(T)\simeq H_{c\text{2}}(0)\left(\frac{Gi}{\pi^2 c_L^4}\right)^{1/3}\!\!\left(1-\frac{T}{T_c}\right)^{2/3}
\label{hm}
\end{equation}   
Taking $c_L=0.17$ and $Gi=10^{-4}-10^{-3}$ in Eq. (\ref{hm}) gives $(Gi/\pi^2c_L^4)^{1/3}\approx 0.23-0.5$, which shows that thermal fluctuations in CaKFe$_4$As$_4$ are not weak, as also characteristic of the majority of FBS which are intermediate between the conventional low-$T_c$ superconductors in which vortex fluctuations are negligible and high-$T_c$ cuprates in which the behavior of vortex matter at 77K is controlled by thermal fluctuations. Yet the width of the critical fluctuation region $T_c-T \lesssim Gi T_c\sim 0.04$ K even at $Gi=10^{-3}$ is still significantly smaller that the observed width of the sharp resistive transition $\Delta T \simeq 0.4$ K at $H=0$ shown in Fig. 1, as well as the width of the step in specific heat in zero field \cite{Meier16}. This suggests that, in addition to thermal fluctuations of the order parameter, the resistive transition at zero field can be broadened by extrinsic factors such as weak materials inhomogeneities in $T_c$. As $H$ increases, the field-induced broadening becomes more pronounced, structural defects and inhomogeneities in $T_c$ affecting both the thermally-activated flux flow resistance \cite{Blatter94} and the vortex melting field \cite{Mikitik03}.   

In conclusion, our magneto-transport measurements of the temperature-dependent anisotropic $H_{c\text{2}}$($T$) of single crystalline CaKFe$_4$As$_4$ up to 63 T show that $H_{c\text{2}}(T)$ is controlled by interplay of orbital and paramangentic effects which cause the anisotropy parameter $\gamma(T)=H_{c\text{2}}^{\text{ab}}(T)/H_{c\text{2}}^{\text{c}}(T)$ to decrease as the temperature decreases. Despite the fact that Ca and K occupy different sites in CaKFe$_4$As$_4$ as opposed to solid solutions in (Ba,K)Fe$_{2}$As$_{2}$, the behavior $H_{c\text{2}}$($T$) turns out to be similar to that of the optimal doped (Ba,K)Fe$_2$As$_2$\cite{Altarawneh08,Tarantini11}. Measurements at lower temperature with higher fields will be needed to reveal more details about the physics of CaKFe$_4$As$_4$.

We would like to thank J. Betts, M. Jaime, R. McDonald, B. Ramshaw, M. Chan, U. Kaluarachchi, V. Taufour, R. Prozorov for useful discussions and experimental assistance. W.R.M. was funded by the Gordon and Betty Moore Foundation EPiQS Initiative through Grant GBMF4411. Work done at Ames Laboratory was supported by US Department of Energy, Basic Energy Sciences, Division of Materials Sciences and Engineering under Contract NO. DE-AC02-07CH11358. The NHMFL Pulsed Field Facility is supported by the National Science Foundation, the Department of Energy, and the State of Florida through NSF Cooperative Grant No. DMR-1157490 and by U.S. DOE BES “Science at 100T” project.


\bibliographystyle{apsrev4-1}

%

\end{document}